# LOSSES IN MONOLAYER GRAPHENE NANORESONATORS DUE TO JOULE DISSIPATION CAUSED BY SYNTHETIC ELECTRIC FIELDS AND THE WAYS OF JOULE LOSSES MINIMIZATIONS

Natalie E. Firsova

Institute for Problem of Mechanical Engineering,
the Russian Academy of Sciences, St. Petersburg 199178, Russia, nef2@mail.ru ,

Yuri A. Firsov

A.F.Ioffe Physical-Technical Institute,
the Russian Academy of Sciences, St.Petersburg, Russia, yuafirsov@rambler.ru

**We consider losses in monolayer graphene nanoresonator connected with Joule dissipation (heating) caused by valley currents stipulated by synthetic electric fields . These synthetic electric fields arise in graphene membrane due to periodic in time gauge fields generated there by external periodic electromotive force. This mechanism accounts for essential part (about 40 percents) of losses in graphene nanoresonator and is specific just for graphene.**
    **The ways of the minimization of these Joule dissipation (increase of the quality-factor of the electromechanical system) are discussed. It is explained why one can increase quality-factor by correctly chosen combination of strains (by strain engineering ). Besides it is shown that quality-factor can be increased by switching on a magnetic field perpendicular to graphene membrane.**

## Introduction

The recent successful preparation of one-atom layer of carbons, graphene [1,2], gave rise to the development of the $2D$-physics. For instance it has provided the opportunity of theoretical and experimental research of relativistic physics in $2D$-world under existing conditions in laboratories of today. Unusual properties of graphene (such as extremely large bending rigidity etc.) proved to be very useful in great number of applications in different spheres of activity.
    In series of new small-size devices named nanoelectromechanical systems (NEMS) (see [3], [4]) the nanoresonators seem to be especially perspective. At first for the fabrication of nanoresonators such materials were used as piezoelectrics, silicon, metallic nanowires, carbon nanotubes. The best dynamic

characteristics may be achieved as the resonator size and mass scaled down (which is assumed in classical linear elastic Bernoulli-Euler beam theory). Resonance frequency may be essentially increased while the quality factor $Q$ will not become much worse (for instance see [5,6]). This allows the sensitive detection of many physical properties such as quantum state, spin, force, molecular mass. These possibilities opened new investigations in biology : virus, protein, and DNA detection, detection of enzymatic activity etc

New opportunities arise if we come to such material as graphene – one carbon atom layer. For instance, recently a new especially precise method was suggested for mass detection (with zg sensitivity) based on NEM mass spectrometer [7] exploiting the advantage of graphene membranes.

Different modifications of graphene nanoresonators were studied, for instance in [8-10]. It was shown that the damping rate increases linearly with resonance frequency. Different kinds of loss mechanisms are discussed in [10-13]. Some of them are common to all experimental setups: attachment losses, thermoelastic dissipation etc. The others depend on actuation scheme, for instance, magnetomotive actuation scheme, capacitive coupling etc. surface-relative losses usually can be modeled by distribution of effective two-level systems. All this possibilities were considered in details in [12]. Authors of [12] pointed out that in graphene nanoresonators dissipation is dominated by electrostatically coupled graphene layer and doped metallic backgate, the energy being dissipated by increasing electron-hole excitations and due to interaction of charge fluctuation with lower-energy flexural phonons.

However such approach does not take into account very specific properties of $2D$-systems. The thing is that in graphene significant role play gauge pseudo-magnetic fields [14] created due to spontaneous generation of large-scale stable distortion of $2D$- graphene surface (ripples, wrinkles…) responsible for its high bending rigidity. These fields radically influence on graphene electronic properties, [15] (where it was discovered that "….graphene sheets are not perfectly flat but exhibit intrinsic microscopic roughening…" and also "…the observed corrugations in the third dimension may shed light on subtle reasons behind the stability of $2D$-crystals") and also [16].

Analytical formulae for pseudo-vector potential $\vec{A}$ for monolayer graphene sheet were obtained for the first time by authors of paper [17] (see also review [18]). There exist expectations that these pseudo-magnetic fields can be used for creation of new graphene nanoelectromechanics. Later it was discovered that these gauge fields may be varied by applying of external strains [19-21] (strain engineering).

However, only in [22] it was pointed out that in graphene one should also take into account that so called synthetic electric fields should arise if pseudo-magnetic gauge fields turn to be time-dependent. Having this idea in mind, authors of [22] calculated damping rates for flexural phonons, their dissipation being caused by these electric fields and associated with them currents (Joule heating).

We'll consider synthetic electric fields which inevitably arise during nanoresonator vibrations driven by external electromotive force. We'll also estimate the resonator intrinsic losses (quality factor $Q$) they cause by heating. We'll show that corresponding contribution in $1/Q$ is very essential and leads to rather big Joule losses in graphene nanoresonators.

Of course, the role of synthetic electric fields in other NEMS may be also important.

In the last section of our paper we discuss the methods of graphene nanoresonators Joule losses reduction .

### *The Model*

For monolayer graphene membrane, the equation of surface being $z = h(x, y)$, for any atom the vectors directed to three nearest neighbors have the form (see for instance [17])

$$\vec{u}_1 = a(\sqrt{3}/2, 1/2), \quad \vec{u}_2 = a(-\sqrt{3}/2, 1/2), \quad \vec{u}_3 = a(0, -1),$$

where $a = 2,5 A$ is a distance between nearest neighbors in the lattice; $h = h(x, y)$ is a distance from a point $(x, y)$ in the plane $XOY$ to the membrane.

In paper [17] the following formulae for gauge field vector potential $\vec{A}$ are obtained

$$A_x(\vec{r}) + iA_y(\vec{r}) = -\sum_j \delta t_j(\vec{r}) e^{i\vec{u}_j \vec{K}} = -\frac{\epsilon_1}{2}\sum_j \left[(\vec{u}_j \cdot \nabla)\nabla h\right]^2 e^{i\vec{u}_j \vec{K}} \qquad (1)$$

$$A_x = -\frac{1}{2}A^0\left[(h_{xx})^2 - (h_{yy})^2\right]a^2, \quad A_y = A^0\left[h_{xy}(h_{xx} + h_{yy})\right]a^2, \qquad (2)$$

$$A^0 = 3/4 \cdot \epsilon_1/e \cdot c/V_F \qquad (3)$$

Here $\epsilon_1 = 2,89 ev$, $\vec{K} = a^{-1}(4\pi/3\sqrt{3}, 0)$ is a Dirac point and $t_j$ - exchange integral with the $j$-th nearest neighbor, $j = 1,2,3$, and $A^0$ has the same dimensionality as vector potential. Products of the expressions in square brackets in formulae for $A_x, A_y$ in (2) by $a^2$ are dimensionless, i.e. they are numerical coefficients, their magnitudes being dependent on the deflection depth of the graphene membrane (we take into consideration large-scale deformations such as ripples, wrinkles etc.) and also on the lattice constant value for the current moment of time.

When switching of alternating electromotive field along the $OZ$ axes the vectors $\vec{u}_j$ should get a time depending variation $\Delta\vec{u}_j(t)$, which is proportional to $E_0 \sin \omega t$, i.e. in linear approximation

$$a(t) = a_0 + \Delta a(t), \qquad (4)$$

where $a_0 = 2{,}5A$ is initial value of the parameter "$a$" at $t = 0$ and

$$\Delta a(t) = \eta_1 E_0 \sin \omega t = a_{00} \sin \omega t \tag{5}$$

Here coefficient $\eta_1$ has dimensionality [cm²/v].
Similarly we assume

$$h(x, y, t) = h_0(x, y) + \Delta h(t) \tag{6}$$

$$\Delta h(t) = \eta_2 E_0 \sin \omega t \cdot \cos(\pi x/2L) = h_{00} \sin \omega t \cdot \cos(\pi x/2L), \tag{7}$$

where $z = h_0(x, y)$ is an equation of the initial membrane surface form and $\eta_2$ has the same dimensionality as $\eta_1$. Both of them describe interaction with actuating field on the microscopic level. The coefficients $\eta_1$, $\eta_2$ may generally speaking depend on $x, y$, but it does not influence the main results of our paper.

Last factor in (7) is connected with the clumping of the opposite membrane edges by $x = \pm L$ (doubly clumped).

Remark that as it is shown in [9] for linear approximation to be reasonable the deflection of graphene nanoresonator vibrations should not be more than 1,1nm. As we assume in our calculations below it is equal to 1nm. Therefore our assumption about linearity is quite reasonable. By the way nonlinear problem was investigated as well in a number of works (see [23] and referencies therein) but we do not study here but linear case.

In presence of the external actuating periodic electric field $E_0 \sin \omega t$ the gauge field vector-potential $\vec{A}$ will depend on time, i.e. in monolayer graphene membrane so called synthetic electric field will arise

$$\vec{E}_{syn} = -c^{-1}\vec{A}_t \tag{8}$$

Let $\omega \approx \omega_{res}$, where $\omega_{res}$ is an eigenfrequency of our resonator. Then substituting (2)- (7) into (8), we find

$$\left(\vec{E}_{syn}\right)_x = -c^{-1}(\vec{A}_x)_t = A^0/c \cdot \{[(h_{xx}^2 - h_{yy}^2)(\Delta a)_t + ah_{xx}(\Delta h)_{xxt}]a\}, \tag{9}$$

$$\left(\vec{E}_{syn}\right)_y = -c^{-1}(\vec{A}_y)_t = -A^0/c \cdot \{h_{xy}[2(h_{xx} + h_{yy})(\Delta a)_t + a(\Delta h)_{xxt}]a\} \tag{10}$$

Using (5), (7), we get

$$\left(\vec{E}_{syn}\right)_x = (E_0\eta_2)E^0(\omega) \cdot$$
$$\{[(\eta_1/\eta_2)(h_{xx}^2 - h_{yy}^2) - ah_{xx}(\pi/2L)^2 \cos(\pi x/2L)]a\} \cdot \cos \omega t \tag{11}$$

$$(\vec{E}_{syn})_y = -(E_0\eta_2)E^0(\omega) \cdot$$
$$\{h_{xy}[2(\eta_1/\eta_2)(h_{xx} + h_{yy}) - a(\pi/2L)^2 \cos(\pi x/2L)]a\} \cdot \cos\omega t \qquad (12)$$

$$E^0(\omega) = 3/4 \cdot \epsilon_1/e \cdot \omega/V_F \qquad (13)$$

We can write formulae (11), (12) in the form

$$(\vec{E}_{syn})_x = E^0(\omega)h_{00}I_x \cos\omega t, \quad (\vec{E}_{syn})_y = E^0(\omega)h_{00}I_y \cos\omega t \qquad (14)$$

where $h_{00} = (E_0\eta_2)$ is a resonator oscillation amplitude (deflection) and

$$I_x = \{[(\eta_1/\eta_2)(h_{xx}^2 - h_{yy}^2) - ah_{xx}(\pi/2L)^2 \cos(\pi x/2L)]a\} \qquad (15)$$

$$I_y = \{h_{xy}[2(\eta_1/\eta_2)(h_{xx} + h_{yy}) - a(\pi/2L)^2 \cos(\pi x/2L)]a\} \qquad (16)$$

Remark that dimensionless quantities $I_x, I_y$ do not turn to zero even by zero deflection because of the presence in graphene of such deformations as ripples, wrinkles and so on.

It follows from (14) – (16) that after time averaging we have

$$(\vec{E}_{syn})^2 = (\vec{E}_{syn})_x^2 + (\vec{E}_{syn})_y^2 = (E^0(\omega))^2 h_{00}^2 (I_x^2 + I_y^2)/2 \qquad (17)$$

Joule losses $\Delta E_J$ for the period $T = 2\pi/\omega$, (similar to the way it was done in [22] in the problem of obtaining damping rate for flexural phonons, (see (14) in [22]) can be written in the form

$$\Delta E_J \approx 2\pi(\vec{E}_{syn})^2 \operatorname{Re}\sigma(\omega) L_x L_y/\omega \qquad (18)$$

where $L_x, L_y$ – membrane sizes. It is shown in [22] that in graphene twodimensional conductivity $\sigma$ does not (or weakly) depend on activating field frequency. So we find

$$\Delta E_J \approx \pi(E^0(\omega))^2 h_{00}^2 (I_x^2 + I_y^2) \sigma L_x L_y/\omega \qquad (19)$$

Remark that in (18), (19) we took into consideration only chargeless ($div\vec{E}_{syn} = div\vec{A}_t = 0$) synthetic electric fields which unlike potential fields are not screened by electrons (see. [22]) and therefore their contribution dominates. Besides in (18), (19) only contribution from one Dirac cone (only one valley, i.e. only one sublattice) is taken into consideration. But as graphene lattice consists of two sublattices (two valleys) we should consider also the field from the second valley. In an ideal case i.e. if there is time-reversal symmetry, [24], these fields have opposite directions and equal magnitudes, and the two valley currents compensate each other. However this question was analysed in [22] where it was shown that two corresponding valley currents do not compensate each other if we take into account intervalley Coulomb drag effect and intervalley scattering on short range impurities.

From formulae (13), (19) we see that the damping rate linearly depends on frequency. It is interesting that in nanoresonators on the basis of carbon nanotubes the dissipation mechanism connected with electron tunneling through vibrating nanotube also gives damping rate linearly depending on frequency, [13].

General losses include different nature parts,

$$Q^{-1} = Q_0^{-1} + Q_J^{-1} \tag{20}$$

Here $Q_0^{-1}$ is connected with dissipation mechanisms studied earlier by other authors (see for instance [8- 13]), and $Q_J^{-1}$ was at first considered and analysed in the present paper.

We introduce quality factor $Q_J$, connected with Joule losses as follows

$$Q_J^{-1} = \Delta E_J / E_{total} \tag{21}$$

Here $\Delta E_J$ was found in (19), and the total energy is defined as follows

$$E_{total} = N \cdot m_{at} \cdot \omega^2 \cdot h_{00}^2, \qquad N = L_x L_y / (a^2 3\sqrt{3}/2),$$

where $N$ is a number of atoms in graphene membrane, $m_{at}$ –is one atom mass and $h_{00}$ - membrane oscillation amplitude. So we obtain

$$Q_J^{-1} = \pi \frac{3\sqrt{3}}{2} \cdot \frac{\left(E^0(\omega)\right)^2 \cdot \sigma \cdot [a^2(I_x^2 + I_y^2)]}{\omega^3 m_{at}} \tag{22}$$

**Joule losses estimation and the ways of their minimization**

Let us estimate the value of Joule losses found in formula (22) and compare the calculated value with experimental data. We consider graphene nanoresonator with frequency $\omega_{res} \approx 130 MHz$ investigated in paper [9]. As for the case $m_{at} = 12 \cdot 1{,}67 \cdot 10^{-24}$, we have $m_{at} \cdot \omega^3 \approx 42 \; [g/s^3]$.

From formula (13) we get

$$E^0(\omega) = 3/4 \cdot \epsilon_1/e \cdot \omega/V_F \approx$$

$$\approx 3/4 \cdot 3 \cdot 1{,}3/3 \; volt/cm = 3{,}9/4 \cdot 1/300 \; CGSE \tag{23}$$

The conductivity for our case was not written in [9] for graphene sample mentioned above. So we take it from another paper ([25]) where the parameters of experiment are close to the ones in [9]. From paper [25] for concentration value $n = 2{,}5 10^{11} [cm^{-2}]$ we find in Fig.1 that $\sigma \approx 1{,}2 \cdot 10^9 \ [cm/s]$ (for good quality of the sample).

Estimate now the factor $a^2(I_x^2 + I_y^2)$ in (22). In [9] it is demonstrated that membrane oscillation critical amplitude after which nonlinearity appears is equal to $1{,}5 nm$. We assume it to be $h_{00} \approx 1 nm$. It is naturally to think that $\Delta a / a \approx h_{00}/h \approx 0{,}1$, i.e.

$$\eta_1/\eta_2 \approx \Delta a / h_{00} \approx (\Delta a / a) \cdot (a / h_{00}) \approx 2{,}5 \cdot 10^{-2}$$

Estimate the first term in the expression for $a^2(I_x^2 + I_y^2)$, using formulae (15), (16). Taking into consideration that graphene membrane surface has corrugations and assuming for simplicity the deformation height (depth) and the basis (lengh, width) to have close sizes we find

$$a^2 \cdot I_x^2 \approx (6{,}25 \cdot 10^{-4} a^4 \cdot (h_{xx}^2)^2 + \cdots) \approx (6{,}25/81) \cdot 10^{-8}$$

When estimating we assumed deformation radius to be $\delta_x \approx 15 nm$ and $\delta h / \delta_x \approx 2$. Other terms in formula for $a^2(I_x^2 + I_y^2)$ can be estimated similarly. Therefore we obtain

$$a^2 (I_x^2 + I_y^2) \approx 0{,}7 \cdot 10^{-8}. \tag{24}$$

Hence and from (22), (23) we find the approximate theoretical numerical value for Joule losses in the sample mentioned above

$$Q_J^{-1} = \Delta E_J / E_{total} \approx 3 \cdot 10^{-5} \tag{25}$$

As experiment in [9] gives the result $Q \approx 14000$ we see that the Joule losses are responsible for about 40 per cent and our model gives the reasonable magnitude of damping rate.

It is interesting that in paper [10] for the sample with about the same resonance frequency they obtained the quality factor $Q \approx 100\,000$. The measured increasing of the quality factor to our point of view was obtained by authors as they used tension. From our formula (22) it is well seen that in this case the factor $(I_x^2 + I_y^2)$ is decreasing which increases the quality factor, i.e. the mesured increasing of quality factor follows from our theory.

Now consider the question how one can minimize the Joule losses $Q_J^{-1}$. It is clear that the expressions $I_x, I_y$ in (15), (16), and consequently the losses (19) can be reduced by varying the form of the function $h(x, y)$ with the help of strains of different kinds. The fact that one can increase the quality factor by such actions was opened experimentally and it has become a subject of a new special field of activity which was called strain engineering. From the formula (22) the reason of this phenomenon is obvious.

One can decrease Joule losses also by switching on magnetic field perpendicular to graphene membrane plane. Indeed according to [26], in this case

$$\sigma_{xx} \approx \sigma(0)/[1 + (\Omega\tau)^2], \qquad (26)$$

where

$$\sigma(0) = 2e^2 h^{-1} V_F \tau \sqrt{\pi n}, \qquad (27)$$

and for Larmour frequency $\Omega$ we have

$$\Omega = V_F h^{-1} (\pi n)^{-1/2} eH/c. \qquad (28)$$

Here $n$ - electron concentration, $V_F$ - Fermi velocity, $\tau$ - relaxation time.

When $\Omega\tau \gg 1$ the value of $\sigma_{xx}(H)$ strongly decreases.

In paper [26] it is shown that one can decrease $\sigma$ by one order of magnitude with the help of magnetic field about 6 Tesla (see. Fig 1 in [26] and also [27]). This gives us possibility decrease the damping (19) by one order of magnitude.

Remark that the formula (26) was obtained using Boltzman equation and stops to be correct when quantization in magnetic field of Landau type starts. Nevertheless though the form of dependence changes the tendency of decreasing conserves.

Since the graphene membrane surface has corrugations the external magnetic field components parallel to vibrating membrane can arise. These components play the role of magnetomotive force. Hence as it is shown in [28-29] we can get extra damping. But as these components are very small compared to the perpendicular one we need not take it into account.

*Conclusion*

In this paper we considered new dissipation mechanism for graphene nanoresonators i.e. Joule losses caused by synthetic electric fields.

For the linear case (i.e. electromotive alternating force is rather small) the formulae for Joule losses are obtained.

We would like to stress especially that in contrast to major part of papers dedicated to nanoresonators in which phenomenological approach within framework of continuum nonlinear elastic model (see [30] and last review-like paper [23]) was used (nonlinear Duffing oscillator) our results for Joule losses are obtained on the basis of microscopic theory taking into account the specific features of graphene. Though membrane vibration is supposed to be classical but the mechanism of losses in graphene nanoresonator is described within the framework of quantum solid state physics.

Using the obtained for Joule losses formulae we calculated approximately their value. This estimate shows that their contribution to the general dissipation proved to be about 40 percents.

The possible methods of lowering down of Joule losses are as follows
- Application of strain engineering methods to minimize quantities $I_x$, $I_y$
- Switching on the magnetic field perpendicular to the graphene membrane.

Note that synthetic currents we considered in this paper lead not only to Joule losses but cause dissipation due to their interaction with currents arising on gate. We hope to analyse these losses in the next paper.

REFERENCES

1. K.S.Novoselov, A.K.Geim, S.V.Morozov, D.Jiang, Y.Zhang, S.V.Dubonos, I.V.Grigorieva, A.A.Firsov, Electric Field Effect in Atomically Thin Carbon Films, Science **306**, 666 (2004). K.S.Novoselov, A.K.Geim, S.V.Morozov, D.Jiang, M.I.Katsenelson, I.V.Grigorieva, S.V.Dubonos, A.A.Firsov, Two-dimensional gas of massless Dirac fermions in graphene, Nature (London) **438**, 197 (2005).
2. Y.Zhang et al, Experimental observation of quantum Hall Effect and Berry's phase in Graphene, Nature (London) **438**, 201 (2005).
3. Andrew N. Cleland Foundation of Nanomechanics, Springer, Berlin, 2003.
4. K.L.Ekinci, M.L.Roukes Nanoelectromechanical systems, Review of scientific instruments **76**, 061101 (2005).
5. A.Gaidarzhy, High quality gigahertz frequencies in nanomechanical diamond resonators, cond-mat 0710. 2613.
6. Kilho Eom, Harold S. Park, Dae Sung Yoon, Taeyun Kwon Nanomechanical Resonators and their Applications in Biological/Chemical Detection: nanomechanics Principles, cond-mat 1105. 1785, To be published in Physics Reports 2011.
7. J.Atalaya, J.M.Kinaret, A.Isacsson Nanomechanical Mass Mesurement using Nonlinear Reponse of a Graphene Membrane, EPL **91**, 48001 (2010), cond-mat/0911.0953 v2.
8. J.Scott Bunch et al, Electromechanical Resonators from Graphene Sheets, Science **315**, 490-493, (2007).
9. Ch.Chen,Sumi, Rosenblatt, Kirill.I.Bolotin et al, Performance of monolayer graphene nanomechanical resonators with electrical readout Nature Nanotechnology,.**4**, 861 (2009).

10. A.Eichler, J.Moser et al, Nonlinear damping in mechanical resonators based on graphene and carbon nanotubes cond-mat / 1103.1788
11  V. Sazonova et al, A tunable carbon nanotube electromechanical oscillator. Nature **431**, 284 (2004).
12  C, Seoanez, F.Guinea, A.H.Castro Neto, Dissipation in graphene and nanotube resonators, Phys. Rev.**B76**, 125427 (2007).
13. Benjamen Lassagne, Yury Tarakanov, Jary Kinaret, David Garcia Sanchez, Adrian Bachtold, Coupling Mechanics To Charge Transport in Carbon Nanotube Mechanical Resonators, Science**. 325,** 1107 (2009).
14. S.V. Morozov, K.S.Novoselov, M.I.Katsnelson, F.Schedin, D.Jiang, A.K.Geim, Strong suppression of weak localization in grapheme, Phys.Rev.Lett. **97**, 016801 (2006).
15. Jannic C. Meyer, A.K.Geim, M.I.Katsnelson, K.S.Novoselov, T.J.Booth,S.Roth The structure of suspended graphene, Nature **446**, 60-63 (2007), cond-mat 070, 1379.
16. A.Fasolino, J.H.Los and M.I.Katsnelson, Intrinsic ripples in graphene, Nature Materials **6**, 858-861 (2007).
17. Eun-Ah Kim, and A.N.Castro Neto, Graphene as an electronic membrane. Phys. Rev. Letters, **84**, 57007 (2008) .
18. M.A.H.Vozmediano, M.I.Katsnelson, F.Guinea, Gauge fields in graphene, Physics Reports **496,** 109 (2010).
19. V.P.Pereira, A.N.Castro Neto, All-graphene integrated cirquits via strain engineering Phys. Rev. Lett. **103**, 046801 (2009).
20. Tony Low, F.Guinea, Strain-induced pseudo-magnetic field for noval electronics. Nano Lett **10**, 3551 (2010).
21. F.Guinea, M.I.Katsnelson, A.K.Geim, Energy gaps, topological insulator state and zero-field quantum Hall effect in graphene by strain engineering, Nature Physics **6,** 30-33 (2010), cond-mat/0909. 1787.
22. Felix von Oppen, Francisko Guinea and Eros Martini, Synthetic electric fields and phonon damping in carbon nanotubes and graphene. Phys.Rev. **80B**, 075420 (2009).
23. J.Moser, A. Eichler, et al, Dissipative and conservative nonlinearity in carbon nanotube and graphene mechanical resonators, arXiv. 1110.1234 v1 [cond-mat. Mes-nan] 6 Oct.2011.
24. A.F.Morpurgo and F.Guinea, Intervalley scattering,long-range disorder,and effective time reversal symmetry breaking in graphene. Phys.Rev.Lett. **97**, 196804 (2006).
25 . Kirill.I.Bolotin et al., Temperature dependent transport in suspended graphene. .Phys.Rev.Lett.**101**, 096802 (2008).
26. Rakesh P.Tivari and D.Stroud, Model for the magnetoresistance and Hall coefficient of inhomogeneous graphene, Phys. Rev. **B79**, 165408 (2009).
27. Kirill I.Bolotin et al., Ultrahigh electron mobility in suspended graphene, Solid State Communication **146,** 351 (2008).

28. A.N.Cleland, M.L.Roukes External control of dissipation in a nanometer-scale radiofrequency mechanical resonator Sensors and Actuators **72**, 256-261 (1999).

29. X.L.Feng and al, Dissipation in single-crystal 3c-sic ultra-high frequency nanomechanical resonators, Nano-Lett. 7, 1953 (2007).

30. L.D.Landau, and E.M. Lifshitsh Theory of Elasticity (Butter-Heinemann and Oxford, 1986), 3$^{rd}$ ed.